\documentclass{ws-procs9x6}
\usepackage{epstopdf}
\begin{document}

\title{Weak Equivalence Principle Test on a Sounding Rocket}

\author{J.D.\ Phillips$^*$,  B.R.\ Patla, E.M.\ Popescu, E.\ Rocco, R.\ Thapa and R.D.\ Reasenberg}

\address{Smithsonian Astrophysical Observatory, Harvard-Smithsonian Center for Astrophysics,\\
60 Garden Street, Cambridge, MA, 02138, U.S.A.\\
$^*$E-mail: jphillips@cfa.harvard.edu}

\author{E.C.\ Lorenzini}

\address{Faculty of Engineering, University of Padova,\\
Italy.\\}

\begin{abstract}
SR-POEM, our principle of equivalence measurement on a sounding rocket, will compare the free fall rate of two substances yielding an uncertainty of $10^{-16}$ in the estimate of $\eta$.  During the past two years, the design concept has matured and we have been working on the required technology, including a laser gauge that is self aligning and able to reach 0.1 pm$/ \sqrt{\text{Hz}}$ for periods up to $40\; \text{s}$.  We describe the status and plans for this project.\\

\end{abstract}

\bodymatter
\section{Introduction}

We are developing a test of the weak equivalence principle (WEP) to be conducted during a 20-minute sounding rocket flight\cite{rdr-jdp-cqg}. It is a Galilean (dropping) test for which the sounding rocket provides both a long free fall time, thus increased signal compared with a ground-based test, and the opportunity for multiple inversions of the apparatus, which allows us to cancel most systematic errors. This is complemented by the high measurement sensitivity of our tracking frequency laser distance gauge (TFG), which is the primary sensor\cite{phillips}. 
For a single pair of test substances, we expect to reach a sensitivity of  $\sigma(\eta) \le 10^{-16}$, a 1000-fold advance over the present state of the art\cite{washington}.

The experiment employs two 0.9 kg test mass assemblies (TMA) in free fall for 40 s per drop. Each TMA is shaped like a dumbbell, having two cubes connected by rods. Each of the four cubes is observed by a TFG; the acceleration of a TMA is taken as the mean of the accelerations derived from its two TFGs. 

During the ascent to 800 km, the TMA are uncaged, captured electrostatically, and electrically neutralized. Then calibrations are performed with the payload on its side so that even in the presence of drag, sensitive measurements can be made along the instrument $z$ axis (WEP measurement direction). The TMA CM are collocated to within about 1 nm along $z$ and 1 $\mu$m in $x$ and $y$. Above 800 km, a series of 8 drops, each lasting 40\;s, and 7 inversions takes place. Then, below 800 km on the way back down, the calibration is repeated. The payload is not recovered.

\section{Systematic error}

The experiment is differential at three levels: First, the TFGs measure distances to the TMA from an accurately comoving instrument. Second, distances to TMA A and B are differenced; also, in estimating acceleration, drifts are suppressed. Third, the payload is inverted between drops. This cancels: (1)\;Gravity from local masses; (2)\;Earth's gradient (the next term does not cancel, but is known and {\it de minimis});\;(3)\;Electrostatic force; (4)\;Outgassing; (5)\;Radiometer effect; (6)\;Thermal radiation; and (7)\;{\bf Some} magnetic terms. To make cancellations (1) and (3) work, the TMA-payload distance is kept constant by a Payload Servo. To make (4)-(6) work, temperature fluctuations are kept below $0.5$\; mK $/\sqrt{\text{Hz}}$ at 0.007\;Hz.

Upon inversion, some components of the magnetic force mimic a WEP signal by remaining unchanged in inertial space. The magnetic force is the product of the TMA magnetic moment and the ambient magnetic field gradient. The most significant component of the latter is due to the Earth's field that penetrates the shield. The gradient is primarily due to the shield. If the shield is symmetric about the plane $z=0$, then in that plane the derivative of the magnetic field with respect to z vanishes. The field gradient affecting SR-POEM depends on  asymmetries of geometry and permeability, and must therefore be tested.

We are now developing the TFG as a tool for NASA precision astronomical missions. We are working on a spaceworthy Semiconductor Laser version (SL-TFG), employing distributed feedback (DFB) lasers emitting at 1560 nm. The accuracy goal for SR-POEM, operating with a resonant cavity measurement interferometer (MI), is 0.1 pm$/ \sqrt{\text{Hz}}$. The TFG presently achieves 2  pm$/ \sqrt{\text{Hz}}$ in a Michelson (non-resonant) MI. Much or all of the required improvement will come from the cavity finesse.

To accommodate the small relative rotations of the TMA and the instrument, we plan to use a system based on the one demonstrated by Sampas and Anderson\cite{sampas} for aligning a measurement beam with a cavity. The advantage of the new system is it functions with a reflected beam.

The TFG has a number of advantages over the classical precision distance gauge, the heterodyne phase gauge. For SR-POEM, the most important are: (1)\;The TFG is free of the cyclic bias of the heterodyne gauge; (2)\;The TFG observable is a radio {\it frequency}, which is much easier to transport and measure stably than the RF {\it phase} observable of the heterodyne gauge; (3)\;The TFG has only one beam, which simplifies the measurement beam launching; (4)\;The TFG can operate in an optically resonant cavity, which increases sensitivity, suppresses misalignment error, and supports an advanced automatic alignment system; and (5)\;The TFG measures absolute distance with a minimum of added complexity.

Coriolis acceleration is one of the most serious error issues. To mitigate it, we require that the payload rotational velocity be 
$<1.3 \times 10^{-4}\;\text{radian\;s}^{-1}$, measured  to 
$<1.4 \times 10^{-7}\;\text{radian\;s}^{-1}$. Further, we set up the TMA so that the difference of their transverse velocity is 
$<5\times 10^{-10}\;\text{m\;s}^{-1}$, measured  to 
$<5\times 10^{-13}\;\text{m\;s}^{-1}$. 
The payload rotational requirements are met by using the star trackers and attitude control system (ACS) thrusters. 
  The TMA velocity requirements are met by using a set of capacitance gauges. The gauges are also used for setup and inversion, and for operating the Payload Servo. During setup and inversion, all six degrees of freedom are measured and the capacitance gauge electrodes are used to apply up to 1100 V over the 4 mm gap to accelerate and decelerate the TMA. 
  During EP measurements, only horizontal measuring fields are applied, and a reduced amplitude is used. Further, the gas jets of the ACS are turned off.

Temperature changes have two types of effects: those in which the measurement of distance is affected directly (e.g. a warp of the TFG plate, which holds one end of each of the four laser gauge cavities), and those in which the measurement is affected indirectly (e.g. a movement of mass near the TMA that causes a change in the local gravity). The direct effect is made small by using ULE for the precision structure, by layered thermal control, and by symmetry of the necessary thermal leaks such as the load-bearing structures. The largest indirect effect comes from masses located near the TMA. Symmetrical design and careful fabrication reduce the differencial acceleration. To further reduce this indirect effect, we employ a Payload Servo to maintain the spacing of the TMA relative to the TFG plate constant to within $\sim 100$\;nm, monitored to an accuracy of $\sim 3$\;nm.

The Payload Servo connects the dual vacuum chamber to the remainder of the payload through a hexapod with actuators (Stewart Platform). Measurements from the TFGs and capacitance gauges determine the spacing between the TMA and their housing. These are fed back to the actuators to keep the spacing constant. Other structures in the vacuum chamber, such as the magnetic shield and chamber walls, are less stably connected to the TMA housing and TFG plate, but are also further from the TMA. The TMA centers of mass are nominally coincident, so the sensitivity of the difference of their accelerations to the positions of these other structures falls off rapidly with distance from the TMA\cite{rdr-jdp-cqg}.

The remaining indirect effects are due to movement of mass far from the TMA. The largest is the transmitter located about 1 m away. During the mission, this heats the payload tube by about 6 K, which expands the payload tube connecting the region of the TMA and the transmitter mass. Based on conservative assumptions and ignoring the cancellation from the inversions, the effect on the WEP estimate is $<10^{-18}$. Thus far, we have not found a thermally-caused problem.

Two sources of thermal perturbation are that the outer rocket skin heats by 130 K during ascent through the atmosphere and, synchronous with payload inversions, the payload is heated radiatively on one side by Earth. The precision instrument, however, hardly sees the external temperature changes. The outer skin is lined with foam insulation and discarded as soon as possible after leaving the atmosphere. The outside of the payload tube and the outside of the vacuum chamber are coated with gold to reduce emissivity. Once above the atmosphere, the space between the chamber and the tube becomes evacuated, and the thermal time constant increases to $1.5\times10^5$\;s. Inside the vacuum chamber, the time constant from wall to the metering structure is $2.8\times10^4$\;s and to the TFG plate it is $1.1\times10^5$\;s.

To restrain the TMA during launch, we must cage them with a force of the order of 1000 N. Soon after motor cutoff, we must release them into free fall gently enough to be captured electrostatically. 
We are experimenting with $n$-docosanethiol, 
 which forms a covalently-bonded monomolecular layer on Au. 
This layer has been shown experimentally to essentially eliminate adhesion between contacting Au surfaces, even after applied force sufficient to cause plastic flow\cite{thomas}. The material and the caging contact would be at the bottom of a recess in the TMA surface, in order to reduce electrostatic disturbances from the contact area.

We have begun laboratory testing of the required technology. In addition to the usual testing, we intend to test much of the required technology and equipment in an aircraft flying ``zero g" parabolic trajectories. Laboratory testing will include: (1)\;We will construct a pair of plates of ULE or other stable glass, one essentially identical to the TFG plate and one to model four TMA. The four measurements will be redundant. In vacuum they will support a TFG test to the required level, $1\times 10^{-13\;}$m$/\sqrt{\text{Hz}}$ at 0.007 Hz. (2)\;We will test the magnetic field gradient present in a shield as similar as possible to that to be used for the flight, including the effect of launch vibration. (3)\;We are addressing the issue of spatial non-uniformity of TMA and housing surface potential in collaboration with J. Cowan 
at the Pacific Northwest National Laboratory, using a scanning Kelvin probe. We plan to test the total electrostatic force by constructing at SAO a torsion pendulum with the help of the Eot-Wash Group at the U. of Washington and the LISA Group at the U. of Trento. (4)\;We have begun discussions with B.\;Buchine regarding making and testing of the $n$-docosanethiol coatings at Harvard's Center for Nanoscale Systems. Once a surface has been created and characterized, we will test its uncaging with a mm-scale indenter that we are now developing. (5)\;TMA magnetic moment will be measured by J.\;Gundlach and collaborators at the University of Washington. Among the approaches to reducing the magnetic moment will be the use of pure materials for the two test substances, and a degaussing procedure similar to that customarily used for ferromagnetic materials. 

Finally we may ask,``Why does SR-POEM work?'' The sounding rocket provides a free fall with low non-gravitational acceleration for $>500$\; s. It also allows for convenient payload inversion, which is a powerful tool for cancelling systematic error. The TFG achieves high precision in a short time. The experiment has multiple layers of passive thermal isolation, providing a thermally-benign environment. And finally, we plan to test many of the systems in an aircraft flying a zero-g trajectory.

\section*{Acknowledgments}
We gratefully acknowledge NASA support under grants NNX07AI11G and NNX08AO04G.

\end{document}